\documentclass[amssymb,superscriptaddress,showpacs,twocolumn,prd]{revtex4}
\usepackage{graphicx}
\def \be{\begin{equation}}
\def \ee{\end{equation}}
\def \bea{\begin{eqnarray}}
\def \eea{\end{eqnarray}}

\newcommand{\beq}[1]{\begin{eqnarray}\label{#1}}
\newcommand{\eeq}{\end{eqnarray}}

\begin{document}
 \pagestyle{plain}

 \title{Does Standard Cosmology Express Cosmological Principle Faithfully?}

 \author{Ding-fang Zeng and Hai-jun Zhao}
 \email{dfzeng@itp.ac.cn} \email{hjzhao@itp.ac.cn}
 \affiliation{Institute of Theoretical
 Physics, Chinese Academy of Science.}
 \begin{abstract}
 In 1+1 dimensional case, Einstein equation cannot give us any information on
 the evolution of the universe because the Einstein tensor of the
 system is identically zero. We study such a 1+1 dimensional
 cosmology and find the metric of it according to cosmological
 principle and special relativity, but the results contradict the
 usual expression of cosmological principle of standard cosmology.
 So we doubt in 1+3 dimensional case, cosmological principle is
 expressed faithfully by standard cosmology.

 \end{abstract}

 \pacs{04.20.Cv, 04.20.Ha, 04.20.Jb}

 \maketitle

 \section{A 1+1 Dimensional Model}
\begin{figure}[h]
 \vspace{-5mm}
 \begin{minipage}[]{0.9\textwidth}\includegraphics[]{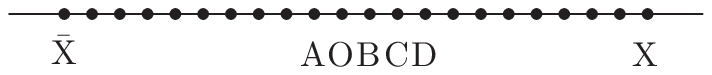}\end{minipage}
 \vspace{3mm}
 \caption{
 One dimensional infinitely long system consists of uniformly
 placed galaxies.
 }
 \label{LorentzContraction}
 \end{figure}

 Consider a one dimensional infinitely long
 system consisting of uniformly placed galaxies, see FIG.\ref{LorentzContraction}. Suppose the system
 is expanding uniformly, i.e., from any galaxies (such as $O$), we will
 see that the two galaxies ($A$ and $B$) mostly nearest to us are recessing away from us
 at equal speeds, and the distances between us and this two
 neighbors are equal.

 In standard cosmology, the scale factor is scale independent,
 i.e., if on the scale of $|OB|$, the scale factor of the system
 is $a(t)$, then on the scale of $|OC|$, the scale factor is also
 $a(t)$. So the physical length of $|OB|$ is half of $|OC|$ and the
 metric describing the system is
 \beq{}
 ds^2=-dt^2+a^2(t)dx^2\label{metricDim1sCDM}
 \eeq
 However, for a one-dimensional gravitation system,
 Einstein equation cannot give us anything about its
 dynamical evolutions, note $G_{\mu\nu}=R_{\mu\nu}-\frac{1}{2}Rg_{\mu\nu}==0$.
 So if we insists eq(\ref{metricDim1sCDM}) is the only correct metric ansaltz
 of the system illustrated in FIG.\ref{LorentzContraction},
 then we have no way to determine the function form of $a(t)$.

 The reason that standard cosmology insists
 eq(\ref{metricDim1sCDM}) is, the density of the system is
 not function of the position co-ordinate $x$. We doubt this
 statement faithfully expresses the requirements of cosmological
 principle. Let us consider the following series
 \beq{}
 &&\hspace{-3mm}v_B=v;\nonumber\\
 &&\hspace{-3mm}v_C=\frac{v+v}{1+v^2};\nonumber\\
 &&\hspace{-3mm}v_D=\frac{v+v_C}{1+v\cdot v_C};\nonumber\\
 &&\textrm{... ...}\nonumber\\
 &&\hspace{-3mm}v_X=\frac{v+v_{X-1}}{1+v\cdot v_{X-1}};
 \label{HubbleSeries}
 \eeq

 \beq{}
 &&\hspace{-3mm}|AB|=2a\nonumber\\
 &&\hspace{-3mm}|OC|=2a\sqrt{1-v_B^2}\nonumber\\
 &&\hspace{-3mm}|BD|=2a\sqrt{1-v_C^2}\nonumber\\
 &&\textrm{... ...}\nonumber\\
 &&\hspace{-3mm}|X^-X^+|=2a\sqrt{1-v_X^2},
 \label{LorentzSeries}
 \eeq
 From which we get
 \beq{}
 v_X=\frac{(1+v)^X-(1-v)^X}{(1+v)^X+(1-v)^X};
 \label{recessionVelocity}
 \eeq
 \beq{}
 |OX|&&\hspace{-3mm}\sim a\sum_{N=0}^{X}{\sqrt{1-v_N^2}}\nonumber\\
 &&\hspace{-3mm}=l\int_0^X dx\sqrt{1-v_x^2}\nonumber\\
 &&\hspace{-3mm}=\frac{4a}{\textrm{ln}\frac{1+v}{1-v}}\left[\textrm{arctg}[(\frac{1+v}{1-v})^{\frac{x}{2}}]-\frac{\pi}{4}\right],
 \label{physicaleDistance}
 \eeq

 In eqs(\ref{HubbleSeries})-(\ref{physicaleDistance}),
 $v$ is the relative recessing speed between two nearest galaxies,
 $a$ is the physical distance between them,
 it can also be considered as the scale factor on the smallest scales,
 \beq{}
 a=v\cdot t,\label{RCosmoSF}
 \eeq
 where $t$ is understood as the time from the system
 being created (the distance between any two galaxies is zero)
 to the epochs we observe it.

 In our models, we will not consider dark energies. But we
 assume that {\bf the relative recessing velocity between any
 two nearest galaxies is a time-independent constance}.
 (One reason for our
 assumption is, if the system is at rest initially, it will not collapse
 at self-gravitations from symmetry analysis, what matters here is parity symmetry. So when the system
 is expanding but cosmological principle is always kept, it will not
 decelerate because the parity symmetry is not broken by expansions)

 Let
 \beq{}
 \sigma=\frac{1}{2}\textrm{ln}\frac{1+v}{1-v}\label{sigmaDefinition}
 \eeq
 we can write down the metric of our one dimensional cosmology in FIG. \ref{LorentzContraction} as
 \beq{}
 ds^2=-dt^2+\frac{4v^2t^2}{(e^{\sigma x}+e^{-\sigma x})^2}dx^2
 \label{metricDim1Cosmo}
 \eeq
 In this metric space, physical co-ordinate $x_{ph}$ is
 related to $x$ by
 \beq{}
 x_{ph}=\frac{2vt}{\sigma}(\textrm{arctg}[e^{\sigma x}]-\frac{\pi}{4}).
 \label{physicaleDistance2}
 \eeq
 Note in eqs(\ref{metricDim1Cosmo}) and
 (\ref{physicaleDistance2}), $x$ is a pure
 number of no-dimensions. Before a length unit is assigned,
 the difference of it has no meaning of any distance lengths.
 But if we let
 \beq{}
 dx_{pr}=dx\cdot vt\label{comovingDistance}
 \eeq
 then $dx_{pr}$ can be naturally interpreted as
 the proper length of line element between points $(t,x)\sim(t,x+dx)$ at time
 $t$. Using co-ordinate $\{t,x_{pr}\}$,
 the metric eq(\ref{metricDim1Cosmo}) can be written in the
 following form
 \beq{}
 ds^2=-dt^2+\frac{4}{(e^{\frac{\sigma x_{pr}}{vt}}+e^{-\frac{\sigma
 x_{pr}}{vt}})^2}dx_{pr}^2\label{metricDim1CosmoSCDMstyle}
 \eeq
 Because we are so deeply affected by Friedmann-Robertson-Walker
 metric and only familiar with only-time-dependent
 (or although both time- and position-dependent in
 the non-flat universes but the two are separated) scale factors,
 we will mostly use
 eq(\ref{metricDim1Cosmo}) in this paper. It is worth noting that
 the $g_{00}$ component of eq(\ref{metricDim1Cosmo}) has different
 dimension from $g_{11}$. Let us put this in mind so that when
 dimension problems appear, correct interpretation can be
 given.

 Some people may ask, why not redefine the
 co-ordinate $x$ so that the $x_{ph}$ has simple
 linear dependence on it? Yes, we can do that way, but we should
 note after the re-definition, the relation between
 $x_{pr}$ and $x$ will change correspondingly, which will
 introduce corresponding complexities, so we
 will not re-define the co-ordinate $x$ at this time.
 Physically, if we re-define $x$ to $\tilde{x}$ so that
 the metric eq(\ref{metricDim1Cosmo}) has the form
 $ds^2=-dt^2+t^2d\tilde{x}^2,\label{coordinateRedefinition}$
 then we should note in equal length of line elements
 $(t,\tilde{x}-\frac{1}{2}d\tilde{x})$ $\sim$
 $(t,\tilde{x}+\frac{1}{2}d\tilde{x})$ and
 $(t,\tilde{x}^\prime-\frac{1}{2}d\tilde{x})$ $\sim$
 $(t,\tilde{x}^\prime+\frac{1}{2}d\tilde{x})$, we will not find
 equal number of galaxies, as long as $|x|\neq|\tilde{x}|$.

 Obviously, eqs(\ref{metricDim1Cosmo}) or
 (\ref{metricDim1CosmoSCDMstyle}) contradicts the standard
 cosmological results eq(\ref{metricDim1sCDM}) remarkablly.
 Standard cosmology insists that cosmological principle requires
 the density of the system is not function of the position
 co-ordinate $x$, so get its metric ansaltz
 eq(\ref{metricDim1sCDM}). While we insists that cosmological does
 not require so, it only requires that on any galaxies, observers
 will measure that his two nearest neighbors are equally far away
 from him and recessing at equal velocities. The density of the
 system can be functions of the position co-ordinate, as long as
 whereever the origin is chosen, the metric function has the same
 form. Although we do not think the generalization from 1+1
 dimension to 1+3 dimension is trivial, we think this is an
 evidence
 that standard cosmology may not express cosmological principle
 faithfully. We will put aside debates at this moment and focus
 on the fact if we generalize the metric eq(\ref{metricDim1Cosmo})
 or (\ref{metricDim1CosmoSCDMstyle}) into three dimensions,
 the prediction is consistent with observations or not. We will
 answer the main criticisms from standard cosmologists in
 \cite{CosmoSDSFext}.

 \section{Generalization To 1+3 Dimensions}

 In generalizing eq(\ref{metricDim1Cosmo}) to three
 dimensional case we have two methods, i.e.,
 \beq{}
 ds^2=-dt^2+\frac{v^2t^2}{\cosh^2{\sigma r}
 }(dr^2+r^2d\theta^2+r^2\textrm{sin}^2\theta d\phi^2)
 \label{metricDim3CosmoPolarSystem}
 \eeq
 or
 \beq{}
 &&\hspace{-3mm}
 ds^2=\nonumber\\
 &&-dt^2+\frac{v^2t^2}{\cosh^2{\sigma x}}dx^2
 +\frac{v^2t^2}{\cosh^2{\sigma y}}dy^2
 +\frac{v^2t^2}{\cosh^2{\sigma z}}dz^2\nonumber\\
 \label{metricDim3CosmoDiscartSystem}
 \eeq
 It is impossible to get eq(\ref{metricDim3CosmoPolarSystem})
 from eq(\ref{metricDim3CosmoDiscartSystem}), because
 eq(\ref{metricDim3CosmoDiscartSystem}) describes a cubic
 lattice system, while eq(\ref{metricDim3CosmoPolarSystem})
 describes a spherical symmetric system. The most remarkable difference between
 eq(\ref{metricDim3CosmoPolarSystem}) and the usual Friedmann-Robertson-Walker metric are,
 (i) the maximum symmetric subspace of metric space eq(\ref{metricDim3CosmoPolarSystem})
 is 2-spheres, while that of standard cosmology
 is homogeneous 3-balls; (ii) eq(\ref{metricDim3CosmoPolarSystem})
 contains no unknown function such as standard cosmology's scale
 factor, the evolution of the universe are completely prescribed by
 one parameter $v$. We will explain these differences in
 \cite{CosmoSDSFext}

 As the first step, let us verify that the metric
 eq(\ref{metricDim3CosmoPolarSystem}) indeed describing a
 homogeneous, isotropic system. Using
 Einstein equation $G_{\mu\nu}=-8\pi GT_{\mu\nu}$ we calculate
 the energy momentum tensor of our cosmology in the following
 \beq{}
 &&\hspace{-3mm}-8\pi GT_{\mu\nu}=\textrm{diag}\nonumber\\
 &&\hspace{-3mm}
 \{
 \frac{-4\sigma(-1+e^{4\sigma r})+\sigma ^2(1-10e^{2\sigma r}+e^{4\sigma r})r-12e^{2\sigma r}rv^2}
   {4e^{2\sigma r}rt^2v^2}\nonumber\\
 &&\hspace{7mm}
 ,\frac{2\sigma(-1+e^{4\sigma r})-\sigma^2(-1+e^{2\sigma r})^2r+4e^{2\sigma r}rv^2}
   {(1+e^{2\sigma r})^2r},\nonumber\\
 &&\hspace{8mm}
 \frac{r(\sigma[-1+e^{4\sigma r}]+4\sigma^2e^{2\sigma r}r+4e^{2\sigma r}rv^2)}
   {(1+e^{2\sigma r})^2},\nonumber\\
 &&\hspace{9mm}
 \frac{r(\sigma[-1+e^{4\sigma r}]+4\sigma^2e^{2\sigma r}r+4e^{2\sigma r}rv^2){\sin^2\theta}}
   {(1+e^{2\sigma r})^2}
 \}
 \nonumber\\
 \label{RCosmoEMT}
 \eeq
 Note in our frame-works, the dimension of $T_{00}$ is different
 from that of $T_{ii}$, because the component of our metric are of different
 dimensions. The same problem will occur on
 the four velocity of an observer, see eq(\ref{RCosmoObeserver}).
 If we use the metric of form
 eq(\ref{metricDim1CosmoSCDMstyle}), this will not be a problem.
 Superficially, our energy momentum tensor is
 position-dependent, which seems to violate cosmological
 principles. This is not the case. Let us calculate
 the energy density and pressures measured by an observer at
 position $(t,r,\theta,\phi)$, whose four velocity can be written as
 \beq{}
 &&\hspace{-3mm}u^{(t,r,\theta,\phi)}=\frac{1}{\sqrt{N}}\{
 1,\ \frac{v_r}{vt},\ 0,\ 0
 \}\cdot
 \nonumber\\
 &&\hspace{0mm}
 \textrm{where }v_r=\frac{e^{\sigma r}-e^{-\sigma r}}{e^{\sigma r}+e^{-\sigma r}},\nonumber\\
 &&\hspace{10mm}N=1-\frac{4v_r^2}{(e^{\sigma r}+e^{-\sigma r})^2}
 .\label{RCosmoObeserver}
 \eeq
 It is very easy to verify $g_{\mu\nu}u^{\mu}u^{\nu}=-1$.
 Measured by this observer, the energy density and pressure are
 respectively
 \beq{}
 &&\hspace{-3mm}8\pi G\rho=\left[T_{\mu\nu}u^{\mu}u^{\nu}\right]_{v\rightarrow0}=9t^{-2},\nonumber\\
 &&\hspace{-3mm}8\pi
 Gp=\left[T_{\mu\nu}(g^{\mu\nu}+u^{\mu}u^{\nu})\right]_{v\rightarrow0}=-9t^{-2}.
 \label{RcosmoEMTmeasure}
 \eeq
 Some people may not understand the limit procedure in
 eq(\ref{RcosmoEMTmeasure}), please see the notations under eq(\ref{SCosmoEMTmeasure}).
 From eqs(\ref{RCosmoEMT})+(\ref{RcosmoEMTmeasure}) we can see
 that, viewing from any point, we can see an isotropic but not in-homogeneous
 universe. The inhomogeneity originates from Lorentz contraction, it is just
 a kinematical effects instead a dynamical one. Obviously, if we can take photos of the universe from
 different places at the same time, we get the same pictures. We
 think this is a faithful expression of cosmological principle.
 While standard cosmology's statements, the energy momentum tensor
 should not depends on the position co-ordinate of the universe,
 does not express cosmological principle faithfully.

 \begin{figure}[h]
 \vspace{5mm}\includegraphics[scale=0.4]{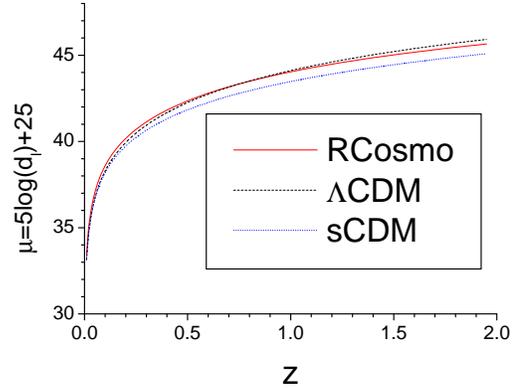}
 \vspace{3mm}
 \caption{
 The luminosity distance v.s. red-shift relation of super-novaes.
 Red(solid) line is the prediction of this paper;
 Black(dot) line is the prediction of $\Lambda$CDM cosmology, in
 which $\Omega_{m0}=0.27$, $\Omega_{\Lambda}=0.73$, $H_0=71\textrm{km/(s}\cdot\textrm{Mpc)}$;
 Blue(dash) line is the prediction of standard CDM cosmology, in
 which $\Omega_{m0}=1.0$,
 $H_0=71\textrm{km/(s}\cdot\textrm{Mpc)}$.
 }
 \label{dlzRelation}
 \end{figure}

 Now let us consider super-novaes in the metric space
 eq(\ref{metricDim3CosmoPolarSystem}).
 We want to calculate their luminosity-distance v.s. red-shift
 relations. Let us follow the same procedures from
 \cite{SWeinberg}, section 14.4, eqs(14.4.11-14).
 Consider a super-novae at position $(t,r,\theta,\phi)$, its
 recessing velocity relative to us is
 \beq{}
 v_r=\frac{e^{\sigma r}-e^{-\sigma r}}{e^{\sigma r}+e^{-\sigma r}};
 \eeq
 so, if the proper frequency of a photon emitted from this
 super-novae is $\omega_{0}$, the frequency measured by us is
 $\omega$, then the red-shift $z$ of this photon satisfy
 \beq{}
 (1+z)\equiv\frac{\omega^{-1}}{\omega_0^{-1}}=e^{\sigma r};
 \eeq
 considering Lorentz dilating, the photons emitted in period $\delta
 t_1$ can only reach us in period $\delta
 t_1e^{\sigma r}$. So we get the luminosity distance v.s.
 red-shift relation as
 \beq{}
 d_l=(1+z)\cdot\frac{2v\cdot H_0^{-1}}{\sigma}[\textrm{arctg}(1+z)-\frac{\pi}{4}]
 \label{dlzRelationForumlae}
 \eeq
 The relation between $\sigma$ and $v$ is given by
 eq(\ref{sigmaDefinition}).
 From best fitting observational results
 of \cite{Riess04}, we get $v=0.79/3000$,
 $H_0=60\textrm{km/(s}\cdot\textrm{Mpc)}$, $\chi^2=303$ (186\textrm{Golden+Silver
 sample}) or $v=0.899/3000$,
 $H_0=$$60\textrm{km/(s}\cdot\textrm{Mpc)}$, $\chi^2=237$ (157\textrm{Golden
 sample}).

 We illustrate the numerical results of
 eq(\ref{dlzRelationForumlae}) in FIG.\ref{dlzRelation}. As
 comparisons, we also depict the predictions of $\Lambda$CDM
 and $s$CDM cosmologies. From the figure we can easily see that
 the prediction of our eq(\ref{dlzRelationForumlae}) is
 very similar to that of $\Lambda$CDM cosmology.
  Because we
 consider special relativity effects on the definition of homogeneity in our theory,
 we call our results eqs(\ref{metricDim1Cosmo}),
 (\ref{metricDim3CosmoPolarSystem}) and (\ref{dlzRelationForumlae}) in this
 paper as Relativity Cosmology.

 \section{Comparing Our Models With Standard Cosmology}

 Now let us return to standard cosmology where we are taught
 that homogeneity and isotropy of the observed universe directly
 leads to the conclusion that Freimann-Robertson-Walker metric is
 the unique metric describing our real universe (we will put perturbations
 and structure formation problems in the future works),
 \beq{}
 ds^2=-dt^2+\frac{a^2(t)}{1-kr^2}dr^2+r^2(d\theta^2+\sin^2\theta
 d\phi^2),\nonumber\\
 \textrm{where,} k=+1,0,-1.\label{SCosmoMetricAnsaltz}
 \eeq
 While in the co-moving co-ordinate
 the energy momentum tensor of the cosmological fluid has
 the form
 \beq{}
 &&\hspace{-3mm}T_{\mu\nu}=\textrm{diag}(\rho,p,p,p)\label{SCosmologyEMT}\nonumber\\
 &&\hspace{7mm}\textrm{if no radiation and/or}\nonumber\\
 &&\hspace{7mm}\textrm{ no dark energy is involved}\nonumber\\
 &&\hspace{4mm}=\textrm{dial}(\rho(t),0,0,0)
 \label{SCosmoEMT}
 \eeq
 \begin{figure}[h]
 \vspace{5mm}\includegraphics[scale=0.45]{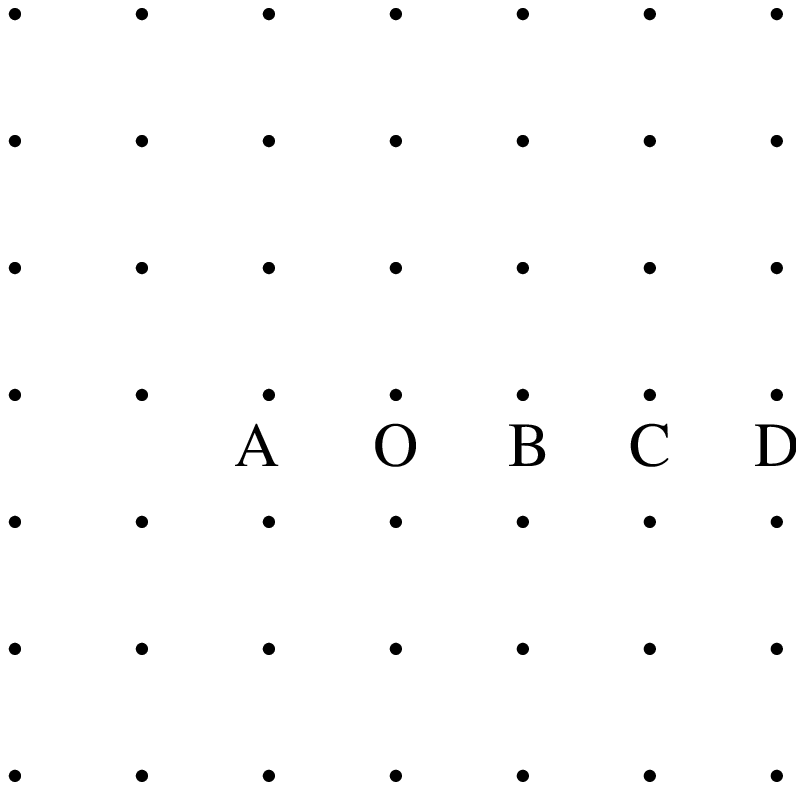}
 \hspace{3mm}\includegraphics[scale=0.47]{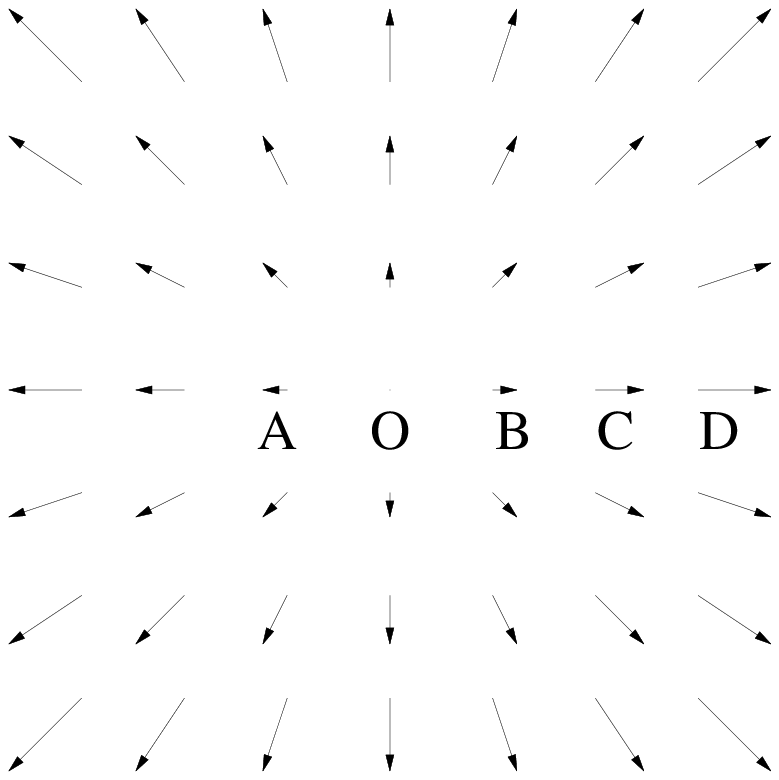}
 \vspace{3mm}
 \caption{
 In a stable universe, homogeneity can be defined
 without any considering of special relativity effects.
 But in an expanding
 universe, the definition of homogeneity is affected by special
 relativity strongly.
 }
 \label{hubbleCurrent}
 \end{figure}

 For a stable universe, special relativity has no
 effects on the homogeneity definition of it,
 please see the left part of FIG. \ref{hubbleCurrent}.
 But for an expanding universe, its homogeneity definition
 is affected by special relativity strongly.
 If we neglect special relativity effects, then
 if we were put on a given galaxy such as $O$ and were asked to measure
 the distances between us and the mostly nearest neighbors,
 see the right part of FIG. \ref{hubbleCurrent},
 we get results, say $r$; if we were asked to measure
 distances between us and the next-nearest neighbors
 (should be on the same line with the previous galaxies) we get
 results $2r$, and so on. So the system has translation
 symmetry at a given time, i.e., the maximum symmetric subspace of
 the whole space-time is homogeneous 3-ball.

 If special relativity is considered,
 then when we were put on a given galaxy $O$ and
 were asked to measure the distances between us
 and the nearest neighbors ($B$), we got results, say $r$.
 But if we were asked to
 measure the distances between us and the next-nearest but
 on the same line neighbor
 ($C$), we did not get $2r$, we got a number less than $2r$
 because of Lorentz contraction. In this case the
 maximum symmetric subspace of our physical
 universe is only a 2-sphere instead of a homogeneous 3-ball.

 According to the results of \cite{SWeinberg},
 section 13.5, the
 general metric describing a space-time with maximum
 symmetric subspace of $S_2$ can only be reduced to
 \beq{}
 ds^2=-dt^2+U(t,r)dr^2+V(t,r)(d\theta^2+\textrm{sin}^2\theta d\phi^2)
 \label{AntiSCosmoMetricAnsaltz}
 \eeq
 instead of eq(\ref{SCosmoMetricAnsaltz}). Just for the
 same reason, we can only write the
 energy momentum tensor describing our real universe as
 \beq{}
 &&\hspace{-3mm}T_{\mu\nu}=\rho u_\mu
 u_\nu+p(g_{\mu\nu}+u_{\mu}u_{\nu})\nonumber\\
 &&\hspace{-3mm}\textrm{where }u^{(t,r,\theta,\phi)}\propto(1,v_r,0,0)
 \label{AntiSCosmoEMT}
 \eeq
 instead of eq(\ref{SCosmoEMT}).
 If generalize our metric eq(\ref{metricDim1CosmoSCDMstyle}) into
 three dimensions it will just has the form of
 eq(\ref{AntiSCosmoMetricAnsaltz}), while the appropriate energy
 momentum tensor eq(\ref{RCosmoEMT}), will also have the form of
 eq(\ref{AntiSCosmoEMT}) correspondingly.

 On the contrary, standard cosmology does not consider
 special relativity when define homogeneities,
 which will introduce problems to it.
 We provide one in the following.
 Starting from metric (\ref{SCosmoMetricAnsaltz}), let $k=0$
 for the moment, using Einstein equation we calculate the energy
 momentum tensor
 \beq{}
 &&\hspace{-3mm}8\pi
 GT_{\mu\nu}=H^2(t)\cdot
 \textrm{Dial.}\{-3,A(t),A(t)r^2,A(t)r^2\sin^2\theta\},
 \nonumber\\
 &&\hspace{20mm}\textrm{where\ }A(t)=a^2(t)+\frac{a^3(t)a^{\prime\prime}}{a^{\prime
 2}}.
 \eeq
 Note $T_{00}$ only depending on $t$ does not mean observers on
 different places will get the equal energy densities in measures.
 It only means
 that energy density measured by observers on the origin of the co-ordinate
 is position independent. To calculate the energy density measured
 by observers on different places, we have to
 consider observers on general positions $(t,r,\theta,\phi)$, whose four velocity
 are
 \beq{}
 u^\mu=\frac{1}{\sqrt{1-[a(t)\dot{a}(t)r]^2}}\{1,\dot{a}(t)r,0,0\}
 \eeq
 The energy density and pressure measured by these observers are
 respectively
 \beq{}
 8\pi G\rho&&\hspace{-3mm}=8\pi GT_{\mu\nu}u^\mu
 u^\nu\nonumber\\
 &&\hspace{-3mm}=\frac{H^2(t)}{1-[a(t)\dot{a}(t)r]^2}(-3+A(t)\dot{a}^2(t)r^2)
 \nonumber\\
 8\pi Gp&&\hspace{-3mm}=8\pi GT_{\mu\nu}(g^{\mu\nu}+u^\mu
 u^\nu)\nonumber\\
 &&\hspace{-3mm}=\frac{5r^2a^2\dot{a}^4-6a\ddot{a}+\dot{a}^2(-3+4r^2{a}^3\ddot{a})}
  {{a}^2(-1+r^2{a}^2\dot{a}^2) }
  \label{SCosmoEMTmeasure}
 \eeq
 Obviously, without a limiting procedure like that
 in eq(\ref{RcosmoEMTmeasure}), eq(\ref{SCosmoEMTmeasure}) will
 tell us that both the energy
 density and pressure measured by
 observers at different places are not the
 same. However, if we take the limiting $\dot{a}\rightarrow0$,
 the energy density of the cosmological fluid will
 become zero. We think this is a problem of standard
 cosmology. But our cosmological models in
 eqs(\ref{metricDim3CosmoPolarSystem})+(\ref{RCosmoEMT})+(\ref{RcosmoEMTmeasure})
 does not have this problem.

 \section{Conclusions}

 We express our suspicions that standard
 cosmology expresses cosmological principle
 faithfully. In 1+1 dimension case, we prove that the background
 metric of the universe is not Friedmann-Robertson-Walker type.
 We then generalize
 the 1+1 dimensional results into 1+3 dimensional case and explain the
 observed luminosity-distance v.s. red-shift relations of
 super-novaes naturally without introducing any concepts of dark
 energies. We will answer the criticisms from standard
 cosmologists in another extended version of this paper,
 \cite{CosmoSDSFext}.

 Of course, the observed luminosity-distance v.s. red-shift
 relations of super-novaes is not the only evidence of dark energies,
 see\cite{Riess98, Perlmutter98, Knop03, Tonry03, Riess04, WMAP03,
 Tegmark04} for experimental works and \cite{Quintessence1, Quintessence2, Phantom, Phantom2, backReaction}
 for theoretical ones. We will study
 the perturbations of eqs(\ref{metricDim3CosmoPolarSystem})+(\ref{RCosmoEMT})
 and structure formation problems in the future.
 The original ideal of this paper is also expressed in
 \cite{Assumption}.

 \begin{center}
 {\bf Acknowledgements\\}
 \end{center}
 When we finish the first version of this paper, we send it to
 professor E. Witten, G. 't Hooft, P. J. Steinhardt, A. Linde, E.
 Kolb and S. Dodelson and ask them to give us some comments or
 criticisms, some of them accepted our asking and give us serious
 comments, we thank them for their comments or
 criticisms or encouragements very much.

\end{document}